\documentclass[11pt,twoside]{article}
\usepackage{asp2014}
\usepackage{pdflscape}
\usepackage{tgschola}
\usepackage{hyperref}
\hypersetup{colorlinks,breaklinks,
            urlcolor=[rgb]{0,0,1.0},
            linkcolor=[rgb]{0,0,1.0}, 
            citecolor=[rgb]{0,0,1.0}} 

\aspSuppressVolSlug
\resetcounters

\newcommand{\teff}{${T}_{\mathrm{eff}}$}
\newcommand{\logg}{$\log{g}$}
\newcommand{\msun}{$M_{\odot}$}

\newcommand{\rsun}{$R_{\odot}$}
\newcommand{\kms}{km s$^{-1}$}

\begin{document}

\title{sdA in SDSS DR12 are Overwhelmingly Not Extremely Low-Mass (ELM) White Dwarfs }
\author{J.~J.~Hermes$^{1,2}$, B.~T.~G\"ansicke$^3$, and Elm\'e~Breedt$^3$
\vspace{2mm}
\affil{$^1$Department of Physics and Astronomy, University of North Carolina, Chapel Hill, NC\,-\,27599-3255, USA; \email{jjhermes@unc.edu}}
\affil{$^2$Hubble Fellow}
\affil{$^3$Department of Physics, University of Warwick, Coventry\,-\,CV4~7AL, UK}}

\paperauthor{J. J. Hermes}{jjhermes@unc.edu}{}{University of North Carolina, Chapel Hill}{Department of Physics and Astronomy}{Chapel Hill}{NC}{27599-3255}{USA}
\paperauthor{B. T. G\"ansicke}{boris.gaensicke@warwick.ac.uk}{}{University of Warwick}{Department of Physics}{Coventry}{Warwickshire}{CV4 7AL}{UK}
\paperauthor{Elme Breedt}{E.Breedt@warwick.ac.uk}{}{University of Warwick}{Department of Physics}{Coventry}{Warwickshire}{CV4 7AL}{UK}

\begin{abstract}
In a search for new white dwarfs in DR12 of the Sloan Digital Sky Survey, \citet{2016MNRAS.455.3413K} found atmospheric parameters for thousands of objects with effective temperatures below 20,000 K and surface gravities between 5.5 < \logg\ < 6.5. They classified these objects as cool subdwarfs --- sdA --- and speculated that many may be extremely low-mass (ELM) white dwarfs (helium-core white dwarfs with masses below 0.3 \msun). We present evidence --- using radial velocities, photometric colors, and reduced proper motions --- that the vast majority (>99\%) of these objects are unlikely to be ELM white dwarfs. Their true identity remains an interesting question.
\end{abstract}

\section{Introduction: Thousands of new cool subdwarfs from SDSS}

In their search for new white dwarfs in the Sloan Digital Sky Survey (SDSS) Data Release 12 (DR12), \citet{2016MNRAS.455.3413K} uncovered a curious population of 2675 stars, which they spectroscopically classified as sdA (see also proceedings here by \citealt{2016arXiv161005550P}). The sdA have hydrogen-dominated spectra with best-fitting atmospheric parameters of 5.5 < \logg\ < 6.5 and \teff\ < $20{,}000$ K. \citet{2016MNRAS.455.3413K} proposed many of these may be extremely low-mass (ELM) white dwarfs.

ELM white dwarfs with masses <0.3 \msun\ cannot have formed in isolation within a Hubble time and by necessity must have close companions. Many have recently been found by the ELM Survey (see, most recently, \citealt{2016ApJ...818..155B}). The 76 ELM white dwarfs with orbits solved so far in the ELM Survey have a median orbital period of 5.4 hr, median radial-velocity (RV) semi-amplitude of $K_1$ = 207 \kms, and median \logg\ = 6.15 cgs. A significant number of ELM white dwarfs have also been found as companions to millisecond pulsars (see Table 4 in the review of \citealt{2008LRR....11....8L}).

Since ELM white dwarfs are the stripped cores of red giants that fail to begin core helium burning, close binarity is a defining characteristic of all ELM white dwarfs. All binaries from the ELM Survey have $P_{\rm orb}<25.8$ hr and are detached (non-interacting); in most cases, this is only possible if both components are compact objects. Using eclipses and ellipsoidal variations, radius constraints on these ELM white dwarfs are consistent with objects having stellar radii less than 0.2 \rsun\ (e.g., \citealt{2014ApJ...792...39H,2014ApJ...794...35G,2016arXiv161206390B}).

\section{Comparing known ELM white dwarfs to sdA}

All SDSS spectroscopic observations are composed of subspectra, typically 3 consecutive 15-min exposures. These data have been exploited since DR6 to search for short-period post-common-envelope binaries (e.g., \citealt{2011A&A...536A..43N}). Using a code to search for RV variability using SDSS subspectra \citep{2017arXiv170205117B}, we have fitted the sdA and the ELM white dwarfs with solved binary paramaters \citep{2016ApJ...818..155B}. The distribution of maximal RVs deduced for these objects are shown in Figure~\ref{fig1maxv}.

\articlefigure{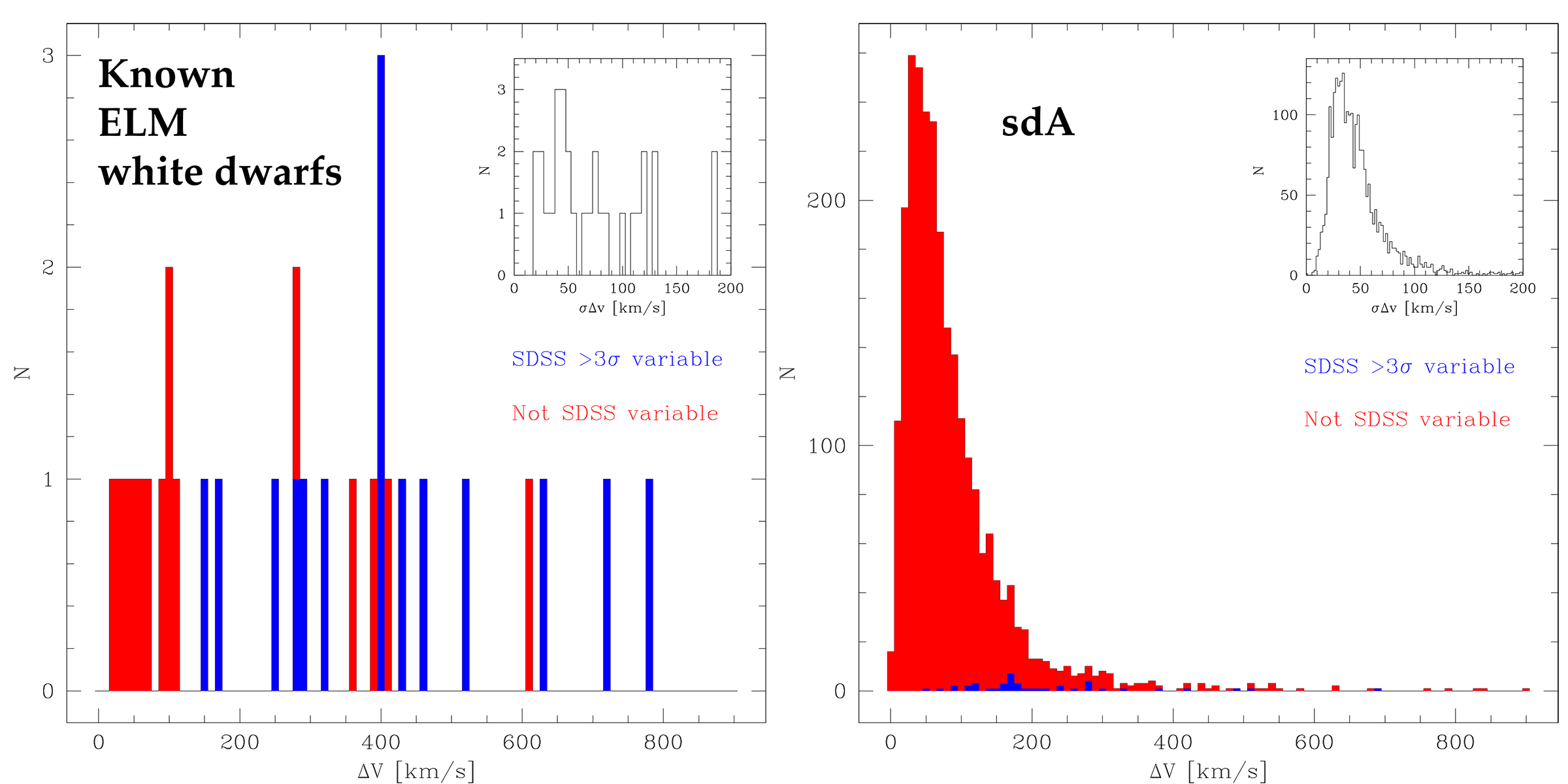}{fig1maxv}{ \emph{Left:} The maximal velocity change among the SDSS subspectra, $\Delta V$, as well as the median velocity uncertainty for the 30 solved ELM white dwarfs with SDSS subspectra. 15 of the 30 (50\%) show significant ($>$3$\sigma$) RV variability, which we mark in blue. Further details for each individual system can be found in Table~\ref{tab2}. \emph{Right:} SDSS subspectra variability of the sdA objects from DR12 proposed by \citet{2016MNRAS.455.3413K}. Just 31 of the 2518 with useable subspectra (1.2\%) show $>$3$\sigma$ RV variability, and are outlined in Table~\ref{tab1}. However, in only three cases (0.1\%) is that variability evident in a single night of subspectra; all 15 of the $>$3$\sigma$ RV variable ELM white dwarfs vary on the timescale of <1 hr.}

We show that half of all known ELM white dwarfs (15 of 30) show $>$3$\sigma$ variability in the SDSS subspectra. The orbital parameters for the $>$3$\sigma$ variables all have $P_{\rm orb}$ < 25.8 hr (median 4.5 hr) and $K_1$ > 55 \kms\ (median 234 \kms). Those without significant variability either have $K_1$ < 130 \kms, $P_{\rm orb}$ < 1 hr, or $g$ > 19.5 mag. The ultracompact 12.75-min ELM binary J0651+2844 \citep{2011ApJ...737L..23B} is one of the 30 with subspectra but is not significantly detected, since each 15-min subspectra covers more than a full orbit. All 15 objects with $>$3$\sigma$ variability show that variability within a single night, and 13 of the 15 show maximal RV variations of $>200$ \kms.

In contrast, just 31 of the 2518 sdA with useable subspectra show $>$3$\sigma$ RV variability; the sdA appear drawn from a significantly different population than the ELM white dwarfs when it comes to short-period RV variability. Only three sdA show varibility in a single night of subspectra, and in one of those three cases the sdA has a clear dM companion: SDSSJ084303.34 +442503.1 ($g$=20.7 mag). Only nine of the 31 objects with $>$3$\sigma$ variability show maximal RV variations of $>200$ \kms. Details of each variable sdA can be found in Table~\ref{tab1}, and can be compared to the 30 known ELM white dwarfs with SDSS subspectra in Table~\ref{tab2}.

Furthermore, we can compare both the known ELM white dwarfs and the sdA in color-color space. In Figure~\ref{elmcolor} we show photometry for the 30 known ELM white dwarfs with SDSS subspectra, which are all single-lined spectroscopic binaries with companions contributing $<5$\% to the optical flux. On the other hand, photometry of the sdA shown in Figure~\ref{sdacolor} show considerably more scatter, especially in the $r-i$ and $i-z$ planes. 

\articlefigure{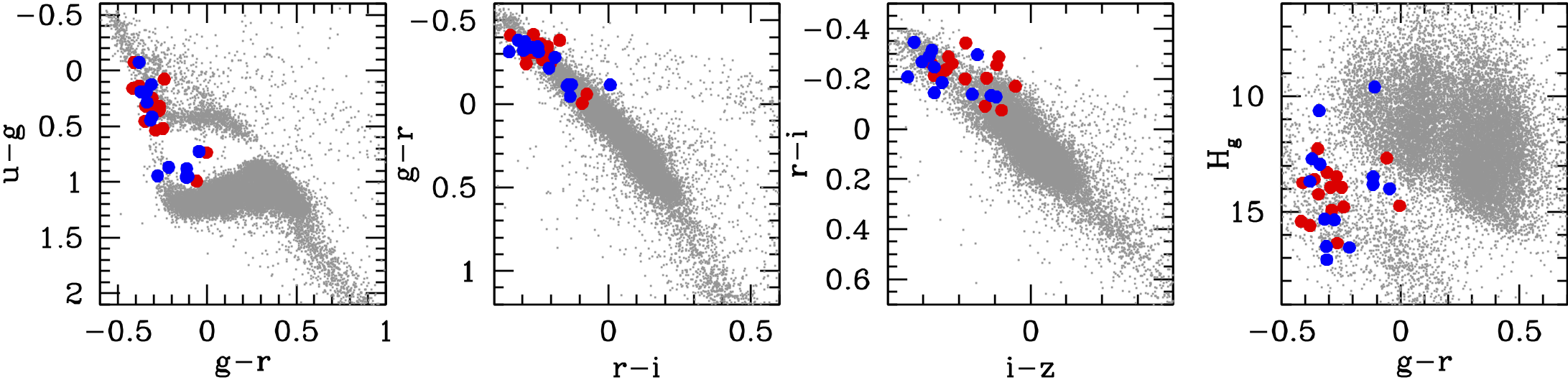}{elmcolor}{Color-color plots show SDSS photometry for the 30 known ELM white dwarfs with SDSS subspectra. Those marked in blue show $>$3$\sigma$ RV variability. Most informative are the $u$-$g$, $g$-$r$ colors, which have been exclusively used in the ELM Survey to select ELM white dwarfs (e.g., \citealt{2016ApJ...818..155B}). ELM white dwarfs still have small radii (more comparable to Neptune than Earth), meaning they are relatively close: We show reduced proper motions in the right panel, following \citet{2015MNRAS.448.2260G}.}

\articlefigure{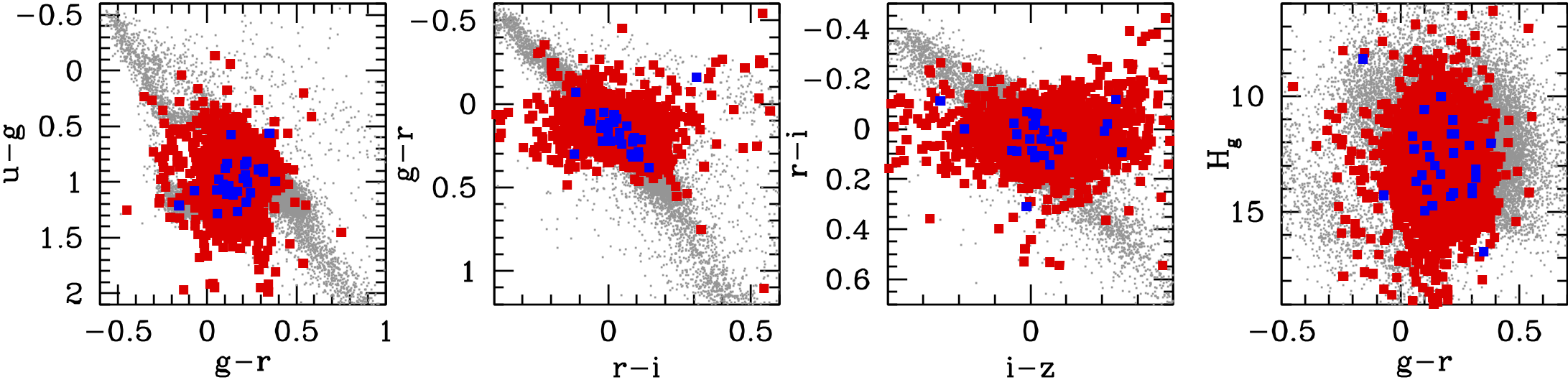}{sdacolor}{Color-color plots show SDSS photometry for the sdA from SDSS DR12 \citet{2016MNRAS.455.3413K}. As above, those marked in blue show $>$3$\sigma$ RV variability. sdA do not appear to follow a distinct track in any color-color plot, but rather show significant scatter. Some sdA could be merger remnants, but their low reduced proper motions makes that unlikely.}

\section{Conclusions: sdA from DR12 are thus overwhelmingly not ELM white dwarfs}

We have shown that, unlike known ELM white dwarfs, the vast majority of sdA in DR12 from \citet{2016MNRAS.455.3413K} do not show short-period RV variability within their SDSS subspectra, and are thus not currently in short-period binaries.

Close binarity is a requirement to form an ELM white dwarf. \citet{2016ApJ...824...46B} find that the majority of low-mass white dwarfs that have been found in the ELM survey should merge within a Hubble time. Thus, the population of sdA objects could be the merged endpoints of previously close ELM white dwarf binaries. However, the majority of sdA do not have significant reduced proper motions, suggesting they have much larger radii than would be expected of a merged ELM white dwarfs. We thus conclude that the sdA from DR12 are overwhelmingly {\em not} extremely low-mass white dwarfs.

While this work is dedicated to establishing that the sdA are not ELM white dwarfs, it is tempting to consider what in fact they are. Given their spread in color-color space, as well as small reduced proper motions, it is possible that some sdA are simply A- and F-stars with low S/N spectra and contamination of the higher-order Balmer lines, artificially increasing the fitted \logg. However, some of these objects may indeed be blue stragglers, low-metallicity field halo stars or other intriguing objects at the final stages of stellar and/or binary evolution, making their overall identity an interesting open question. Distances deduced from GAIA parallaxes should provide sufficiently accurate estimates of the radii of these objects to better illuminate the sdA population before the 21st European White Dwarf Workshop.

Given this potentially large population of stars masquerading as low-surface-gravity white dwarfs, it would be wise to only define an ELM white dwarf as an object with confirmed short-period RV variability, such that it must be a compact object to remain detached from its close binary companion. To date, there are 12 published objects with \logg\ $> 5.0$ discovered by the ELM Survey that do not show RV variability, to limits of roughly 30 \kms\ \citep{2016ApJ...818..155B}; the majority of those 12 non-variable objects have \teff\ $\simeq8000$\,K, likely making them much more akin to the sdA than low-mass white dwarfs.

\acknowledgements Support for this work was provided by NASA through Hubble Fellowship grant \#HST-HF2-51357.001-A, awarded by the Space Telescope Science Institute, which is operated by the Association of Universities for Research in Astronomy, Incorporated, under NASA contract NAS5-26555. The research leading to these results has received funding from the European Research Council under the European Union's Seventh Framework Programme (FP/2007-2013) / ERC Grant Agreement n. 320964 (WDTracer).

%

\begin{landscape}
\centering
\begin{table}
\caption{Objects classified as sdA from \citet{2016MNRAS.455.3413K} with $>$3$\sigma$ RV variability within SDSS subspectra. \label{tab1}}
\begin{tabular}{@{}lccrrrl@{}}
\hline
SDSS ID              & $g$   & SDSS & $\Delta V_{\rm max}$  & \teff\   &  \logg\   & Notes  \\
(J2000.0)            & (mag) & Epochs & (\kms)                & (K)      & (cgs)     &         \\
\hline
J101409.25$+$591753.3 & 19.9 & 3 & 424(198) & 8230(60) & 6.26(0.17) & Low signal-to-noise\\
J084303.34$+$442503.1 & 20.7 & 1 & 285(62) & 11330(360) & 4.36(0.04) & {\bf Variability within single night; sdA+dM}\\
J150552.50$+$202519.3 & 19.8 & 3 & 285(68) & 7040(50) & 5.62(0.20) & Low signal-to-noise\\
J121318.75$+$085644.6 & 19.5 & 1 & 278(65) & 6930(80) & 5.74(0.21) & \\
J094751.94$+$560213.8 & 20.0 & 2 & 261(76) & 8450(60) & 5.78(0.24) & \\
J022740.49$-$002230.1 & 22.4 & 6 & 239(69) & 6900(80) & 5.68(0.23) & Low signal-to-noise\\
J020119.50$-$004935.4 & 17.9 & 5 & 217(34) & 7760(10) & 5.68(0.05) & \\
J153341.94$+$490953.1 & 20.6 & 1 & 211(52) & 11250(280) & 4.47(0.04) & \\
J094637.58$+$444737.4 & 19.5 & 2 & 205(68) & 7450(60) & 5.97(0.16) & \\
J022932.45$-$001427.8 & 18.4 & 23 & 181(37) & 6780(30) & 5.69(0.08) & Possible systemic velocity offsets\\
J105855.99$+$004428.5 & 20.2 & 2 & 181(55) & 7220(90) & 5.63(0.30) & Possible short-period variability\\
J035255.41$-$060336.3 & 19.2 & 3 & 176(49) & 7410(40) & 5.61(0.18) & \\
J011536.38$+$315054.1 & 19.2 & 3 & 172(41) & 6850(50) & 5.54(0.17) & Likely long period\\
J142908.88$+$030125.6 & 19.6 & 2 & 172(51) & 7660(70) & 6.42(0.22) & Low signal-to-noise\\
J090855.18$+$565214.9 & 19.0 & 3 & 169(36) & 7870(20) & 6.11(0.08) & \\
J074422.20$+$461404.5 & 19.1 & 3 & 168(49) & 10320(120) & 6.00(0.00) & Likely long period\\
J150554.06$+$251925.8 & 19.9 & 2 & 167(50) & 7370(60) & 5.80(0.24) & \\
J112837.73$-$000112.5 & 18.4 & 4 & 167(34) & 7900(20) & 5.70(0.07) & Likely long period\\
J074220.62$+$485741.5 & 18.3 & 3 & 164(37) & 7980(60) & 5.85(0.21) & \\
J030305.44$-$003428.9 & 18.8 & 5 & 158(39) & 7710(30) & 6.07(0.09) & \\
J021720.33$+$070935.2 & 18.5 & 2 & 155(29) & 8180(30) & 5.66(0.11) & Likely systemic velocity offset\\
J141632.25$+$523215.2 & 18.6 & 4 & 149(65) & 6670(80) & 5.92(0.25) & \\
J034436.19$+$001857.6 & 19.2 & 8 & 141(33) & 19130(190) & 7.96(0.03) & DA, internally misclassified\\
J113928.25$-$012610.2 & 19.0 & 5 & 119(38) & 8170(30) & 5.79(0.08) & \\
J080737.09$+$143314.6 & 18.8 & 1 & 116(26) & 6840(50) & 5.67(0.14) & {\bf Variability within single night}\\
J130315.83$+$105752.3 & 18.9 & 4 & 111(26) & 19750(190) & 5.74(0.02) & \\
J002459.29$+$332727.3 & 19.3 & 3 & 110(29) & 6950(50) & 5.55(0.15) & \\
J073343.80$+$260011.9 & 18.9 & 2 & 92(25) & 7370(40) & 5.55(0.14) & {\bf Variability within single night}\\
J120324.47$+$194835.3 & 17.5 & 8 & 88(21) & 7730(20) & 5.55(0.05) & \\
J080238.51$+$382546.6 & 18.6 & 3 & 71(21) & 6910(70) & 5.60(0.25) & \\
J002921.38$+$182010.6 & 18.8 & 1 & 51(16) & 6830(30) & 5.59(0.11) & \\
\hline
\end{tabular}
\end{table}

\clearpage

\begin{table}
\caption{Results from known ELM white dwarfs with solved orbital parameters \citep{2016ApJ...824...46B}. \label{tab2}}
\begin{tabular}{@{}lccrrrrrr@{}}
\hline
SDSS ID               & $g$   & SDSS & $\Delta V_{\rm max}$  & \teff\   &  \logg\   & $K_1$ & $P_{\rm orb}$ & RV var.  \\
(J2000.0)             & (mag) & Epochs  & (\kms)                & (K)      & (cgs)     & (\kms) & (hr) & from SDSS?     \\
\hline
J010657.40$-$100003.3 & 19.6 & 1 & 779(133) & 16970(260) & 6.10(0.05) & 395.2(3.6) & 0.6516 & {\bf yes, $>$4$\sigma$} \\
J084910.13$+$044528.7 & 19.1 & 1 & 718(108) & 10290(150) & 6.29(0.05) & 366.9(4.7) & 1.8888 & {\bf yes, $>$4$\sigma$} \\
J082511.90$+$115236.4 & 18.6 & 2 & 628(82) & 27180(400) & 6.60(0.04) & 319.4(2.7) & 1.39656 & {\bf yes, $>$4$\sigma$} \\
J105611.02$+$653631.5 & 19.7 & 1 & 614(185) & 21010(360) & 7.10(0.05) & 267.5(7.4) & 1.04424 & no, $<$3$\sigma$ \\
J143633.29$+$501026.7 & 18.2 & 1 & 522(51) & 17370(250) & 6.66(0.04) & 347.4(8.9) & 1.0992 & {\bf yes, $>$4$\sigma$} \\
J123800.09$+$194631.4 & 17.2 & 2 & 464(20) & 14950(240) & 4.89(0.05) & 258.6(2.5) & 5.346 & {\bf yes, $>$4$\sigma$} \\
J092345.59$+$302805.0 & 15.6 & 1 & 427(27) & 18500(290) & 6.88(0.05) & 296.0(3.0) & 1.0788 & {\bf yes, $>$4$\sigma$} \\
J002207.65$-$101423.5 & 19.6 & 4 & 414(121) & 20730(340) & 7.28(0.05) & 145.6(5.6) & 1.91736 & no, $<$3$\sigma$ \\
J011210.24$+$183503.8 & 17.1 & 2 & 405(22) & 10020(140) & 5.76(0.05) & 295.3(2.0) & 3.52752 & {\bf yes, $>$4$\sigma$} \\
J210308.79$-$002748.8 & 18.2 & 1 & 403(43) & 10130(150) & 5.78(0.05) & 281.0(3.2) & 4.87392 & {\bf yes, $>$4$\sigma$} \\
J075519.46$+$480034.4 & 15.9 & 3 & 402(26) & 19520(300) & 7.42(0.05) & 194.5(5.5) & 13.11048 & {\bf yes, $>$4$\sigma$} \\
J065133.33$+$284423.4 & 18.8 & 3 & 392(114) & 16340(260) & 6.81(0.05) & 616.9(5.0) & 0.212557373 & no, $<$3$\sigma$ \\
J100548.09$+$054204.4 & 19.7 & 1 & 357(187) & 16590(260) & 7.38(0.05) & 208.9(6.8) & 7.3344 & no, $<$3$\sigma$ \\
J123316.19$+$160204.6 & 19.8 & 1 & 320(75) & 11700(240) & 5.59(0.07) & 336.0(4.0) & 3.6216 & {\bf yes, $>$3$\sigma$} \\
J105353.89$+$520031.0 & 18.9 & 1 & 287(74) & 16370(240) & 6.54(0.04) & 264.0(2.0) & 1.02144 & {\bf yes, $>$3$\sigma$} \\
J163030.59$+$423305.8 & 19.0 & 1 & 277(119) & 16070(250) & 7.07(0.05) & 295.9(4.9) & 0.66384 & no, $<$3$\sigma$ \\
J100554.05$+$355014.2 & 19.0 & 1 & 276(50) & 10060(140) & 6.02(0.05) & 143.0(2.3) & 4.23648 & {\bf yes, $>$4$\sigma$} \\
J114155.56$+$385003.0 & 19.0 & 1 & 245(69) & 11290(210) & 4.94(0.10) & 265.8(3.5) & 6.22392 & {\bf yes, $>$3$\sigma$} \\
J151826.68$+$065813.2 & 17.5 & 2 & 171(41) & 9940(140) & 6.82(0.04) & 172.0(2.0) & 14.6244 & {\bf yes, $>$4$\sigma$} \\
J100559.10$+$224932.3 & 17.3 & 1 & 154(39) & 11680(310) & 7.32(0.02) & 176.1(1.1) & 2.7843744 & {\bf yes, $>$3$\sigma$} \\
J084523.03$+$162457.5 & 19.7 & 1 & 111(129) & 19620(310) & 7.51(0.05) & 62.2(5.4) & 18.14376 & no, $<$3$\sigma$ \\
J081544.25$+$230904.7 & 17.6 & 3 & 100(47) & 21430(330) & 5.84(0.05) & 131.7(2.6) & 25.76568 & no, $<$3$\sigma$ \\
J002228.44$+$003115.5 & 19.3 & 1 & 95(102) & 20460(310) & 7.58(0.04) & 80.8(1.3) & 11.7924 & no, $<$3$\sigma$ \\
J162542.10$+$363219.1 & 19.3 & 1 & 88(65) & 24700(400) & 6.10(0.05) & 58.4(2.7) & 5.57712 & no, $<$3$\sigma$ \\
J084037.57$+$152704.5 & 19.2 & 1 & 65(43) & 13670(230) & 5.04(0.05) & 84.8(3.1) & 8.268 & no, $<$3$\sigma$ \\
J073032.89$+$170357.0 & 19.7 & 1 & 64(84) & 12030(220) & 6.30(0.05) & 122.8(4.3) & 16.7448 & no, $<$3$\sigma$ \\
J155708.48$+$282336.0 & 17.6 & 1 & 52(54) & 12560(190) & 7.76(0.05) & 131.2(4.2) & 9.72984 & no, $<$3$\sigma$ \\
J123410.36$-$022802.8 & 17.7 & 1 & 41(41) & 17800(260) & 6.61(0.04) & 94.0(2.3) & 2.19432 & no, $<$3$\sigma$ \\
J082212.57$+$275307.4 & 18.2 & 1 & 26(35) & 8980(130) & 6.65(0.05) & 271.1(9.0) & 5.856 & no, $<$3$\sigma$ \\
J110436.70$+$091822.1 & 16.6 & 1 & 19(28) & 16700(260) & 7.61(0.05) & 142.1(6.0) & 13.27656 & no, $<$3$\sigma$ \\
\hline
\end{tabular}
\end{table}

\end{landscape}
\end{document}